\newcommand \bs{\begin{subequations}}
\newcommand \es{\end{subequations}}
\newcommand \bea{\begin{eqnarray}}
\newcommand \eea{\end{eqnarray}}
\newcommand \be{\begin{equation}}
\newcommand \ee{\end{equation}}
\newcommand \nn{\nonumber}

\documentclass[
 aps,
 prl,
 amssymb,showpacs,
 twocolumn,
  ]{revtex4}

\usepackage{graphicx}
\usepackage{amsmath}

\begin{document}

\title{Transport of quantum noise through random media}

\date{\today}
\author{P. Lodahl and A. Lagendijk}

\affiliation{ Complex Photonic Systems, Department of Science and
Technology and MESA+ Research Institute, University of Twente,
P.O. Box 217, 7500 AE Enschede, The Netherlands}


\begin{abstract}
We present an experimental study of the propagation of quantum
noise in a multiple scattering random medium. Both static and
dynamic scattering measurements are performed: the total
transmission of noise is related to the mean free path for
scattering, while the noise frequency correlation function
determines the diffusion constant. The quantum noise observables
are found to scale markedly differently with scattering parameters
compared to classical noise observables. The measurements are
explained with a full quantum model of multiple scattering.
\end{abstract}

 \maketitle

Light propagation in static disordered photonic media is coherent.
The coherence is preserved even after a very large number of
scattering events. Coherent transport of light in a disordered
medium is the basis for applications of wave scattering to enhance
communication capacities \cite{Moustakas00}, for acoustical and
biomedical imaging, as well as for fundamental discoveries of
intensity correlations, enhanced backscattering, and Anderson
localization \cite{Sheng95}. All these phenomena are captured by
classical physics where for light the electromagnetic field is
described by Maxwell's wave equation. In contrast, no experiments
on multiple scattering have been carried out yet in the realm of
quantum optics where a classical description of light is
insufficient and effects of the quantized nature of the
electromagnetic field are encountered. Pioneering theoretical work
includes a study of the propagation of coherent \cite{Patra99} and
squeezed light \cite{Patra00} through a random medium with gain,
and the associated quantum noise limited information capacities in
such random lasers \cite{Tworzydlo}. The photon statistics of a
random laser was recently measured \cite{Cao01}, which confirmed
the expectations for a laser, and is an instructive example of the
independent information that can be extracted by quantum optical
measures. Here we present measurements of the transport of quantum
and classical noise through a passive multiple scattering medium.

Noise is inevitable in all measurements. The fundamental lower
limit is determined by quantum mechanics through Heisenberg's
uncertainty relation, and is referred to as the shot noise limit.
In the particle description of quantum mechanics, the existence of
optical shot noise directly proves that light is quantized
\cite{Henry96}. Shot noise fluctuations scale proportional to the
square root of the average number of photons (particle-like
behavior) in contrast to classical fluctuations that scale
linearly (wave-like behavior). The different scaling allows us to
distinguish quantum noise from classical noise in an optical
experiment. Shot noise is universal to systems consisting of
quantized entities and, e.g., offers independent information about
the conduction of electrons in mesoscopic conductors compared to
standard conductance measurements \cite{Blanter00}.

In the current Letter we investigate the propagation of classical
and quantum intensity noise of light through a multiple
scattering, randomly ordered medium. Two different measurements
are presented in the noise: total transmission and (short range)
frequency correlations. The two measurements provide insight in
static and dynamic transport of quantum noise in a random medium,
respectively.

Multiple scattering of light forms a volume intensity speckle
pattern of bright and dark spots that most conveniently can be
described as a discrete number of conduction channels. We consider
the intensity transmitted from an input channel $a$ to an output
channel $b$, $I_{\omega}^{ab}(t),$ which depends on time $t$ and
the optical frequency $\omega.$ The intensity is expanded as a
mean value $\overline{ I_\omega^{ab}}$ plus a fluctuating part
$\delta { I_\omega^{ab}}(t)$ that describes the quantum noise. In
a total transmission measurement we sum up all output channels,
and the total transmitted intensity is $I_{\omega}^{T}(t) = \sum_b
I_{\omega}^{ab}(t).$ The noise transmission coefficient is defined
as
\be \mathcal{T}_a^N(\Omega) = \frac{ \overline{ \left| \delta
I_\omega^{T}(\Omega)\right|^2 }}{\overline{\left| \delta
I_\omega^{in}(\Omega)\right|^2}} =  \frac{ \overline{ \left|
\sum_b \delta I_\omega^{ab}(\Omega)\right|^2 }}{\overline{\left|
\delta I_\omega^{in}(\Omega)\right|^2}}, \label{TotalTN}\ee
where the bars denote average over measurement time and $\Omega$
is the frequency (Fourier transform of $t$) that accounts for
slowly varying intensity fluctuations of light. $\overline{\left|
\delta I_\omega^{in}(\Omega)\right|^2}$ is the spectral density of
the input noise of the light illuminating the sample through
channel $a$. For a fixed output channel b, we furthermore define
the noise auto-correlation function for a frequency offset $\Delta
\omega$:
\bea &&\left< \left< C_{ab}^N(\Delta \omega,\Omega) \right>
\right>_{\omega} = \label{corrfcts}\\
&& \frac{\left< \left< \overline{ \left| \delta
I_\omega^{ab}(\Omega)\right|^2} \times \overline{ \left| \delta
I_{\omega + \Delta \omega }^{ab}(\Omega)\right|^2} \right> \right>
_{\omega} - \left< \left< \overline{ \left| \delta
I_\omega^{ab}(\Omega)\right|^2} \right> \right>_{\omega}^2
}{\left< \left< \overline{ \left| \delta
I_\omega^{ab}(\Omega)\right|^2} \right> \right>_{\omega}^2},
  \nn \eea
where double brackets $\left< \left< \cdot \cdot  \right>
\right>_{\omega}$ denote ensemble average that in this case is
obtained by averaging over the optical frequency $\omega.$ Such
noise correlation functions are introduced here for the first
time, while substantial efforts have centered on intensity
correlation functions \cite{Genack90}.

The noise spectral density of the transmitted light can be
calculated using a full quantum model for multiple scattering
\cite{Patra99}. We relate the annihilation operator of the output
electric field in channel $b$ ($\hat{a}_{\omega}^{ab}$) to the
input electric field in channel $a$ ($\hat{a}_{\omega}^{a}$)
through the relation
\be \alpha_{\omega}^{ab} + \hat{a}_{\omega}^{ab}(t) =
t_{\omega}^{ab} \left[ \alpha_{\omega}^{a} +
\hat{a}_{\omega}^{a}(t) \right] + \sum_{a' \neq a} t_\omega^{a' b}
\hat{a}_{\omega}^{a'}(t) + \sum_{b'} r_\omega^{b'b}
\hat{a}_{\omega}^{b'}(t), \label{a}\ee
where indices $a'$ and $b'$ label channels on the input and output
side of the multiple scattering medium, respectively.
$\hat{a}_{\omega}^{a'}(t)$ and $\hat{a}_{\omega}^{b'}(t)$ account
for vacuum fluctuations in all open channels while
$t_{\omega}^{a'b}$ and $r_{\omega}^{b'b}$ are electric field
transmission and reflection coefficients. In Eq. (\ref{a}), we
have specified the coherent amplitudes of the input field
$(\alpha_{\omega}^a)$ and the output field
$(\alpha_{\omega}^{ab})$, and for a coherent state the remaining
fluctuations equal vacuum fluctuations, i.e.
$\left<\hat{a}_{\omega}^{a}(t) \right> =
\left<\hat{a}_{\omega}^{b}(t) \right> =
\left<\hat{a}_{\omega}^{ab}(t) \right> = 0$ for all $a$ and $b$.
Since we are concerned with intensity fluctuations at frequencies
$\Omega$ that are very slow compared to the characteristic
frequency for transport and change of phase through the scattering
sample $(D/L^2 \sim 10^{12} \: \mathrm{Hz})$ \cite{Tiggelen99} as
well as the optical frequency $(\omega \sim 10^{15} \:
\mathrm{Hz}),$ it is an excellent approximation to employ a
single-longitudinal frequency $(\omega)$ for the optical field.
Fourier transforming Eq. (\ref{a}) we calculate the spectral
density of the intensity fluctuations $\left| \delta I(\Omega)
\right|^2 = \left<\left( \delta \hat{I}(\Omega)\right)^2 \right>,$
where $\hat{I} = \hat{a}^{\dagger} \hat{a}$ and $\delta
\hat{I}(\Omega) = \left[\alpha^* + \hat{a}^{\dagger}(\Omega)
\right] \left[\alpha + \hat{a}(\Omega) \right] - \left| \alpha
\right|^2$ is a self-adjoint operator. The spectral density is the
quantity measured in the experiment, and for a single output
channel $b$ we obtain \cite{Patra00}
 \bea  &&\overline{\left| \delta I_{\omega}^{ab}(\Omega) \right|^2}
 = \label{deltaI} \\
 && \left| {t_{\omega}^{ab}} \right|^4 \left( \overline { \left|
\delta I_{\omega}^{in}(\Omega) \right|^2 } - \overline{ \left|
\delta I_{\omega}^v(\Omega) \right|^2 } \right) +
\left|{t_{\omega}^{ab}}\right|^2 \overline{ \left| \delta
I_{\omega}^v(\Omega) \right|^2 }, \nn  \eea
where we have defined the vacuum contribution
\be \overline{\left| \delta I_{\omega}^{v}(\Omega) \right|^2}
\equiv \overline{I_{\omega}^{in}} \left<
\hat{a}_{\omega}^{b'}(\Omega)
(\hat{a}_{\omega}^{b'}(\Omega))^{\dagger} \right>,  \ee
that results from beating between the input field in channel $a$
and vacuum fluctuations from each of the vacuum channels $a' \neq
a$ and $b'$ \cite{Henry96}. We note that $ \left<
\hat{a}_{\omega}^{b'}(\Omega)
(\hat{a}_{\omega}^{b'}(\Omega))^{\dagger} \right> = 1.$ Classical
noise (in the following referred to as technical noise) can also
be described with Eq. (\ref{deltaI}) by neglecting all vacuum
contributions. In the case of shot noise (SN), we have $\overline{
\left| \delta I_{\omega}^v(\Omega) \right|^2} = \overline{ \left|
\delta I_{\omega}^{in}(\Omega) \right|^2},$ and consequently
\bs \bea && \mathcal{T}_{ab}^{SN} = \frac{\overline{ \left| \delta
I_{\omega}^{ab}(\Omega) \right|^2_{SN}}}{\overline{ \left| \delta
I_{\omega}^{in}(\Omega) \right|^2}} =
{T_{\omega}^{ab}}, \\
&& \mathcal{T}_{ab}^{TN} = \frac{\overline{ \left| \delta
I_{\omega}^{ab}(\Omega) \right|^2_{TN}}}{\overline{ \left| \delta
I_{\omega}^{in}(\Omega) \right|^2}} ={T_{\omega}^{ab}}^2, \eea
\label{TN-SNonespot} \es
where $T_{\omega}^{ab} = \left|{t_{\omega}^{ab}}\right|^2$ is the
intensity transmission coefficient from channel $a$ to $b$. The
noise of the total transmitted intensity can be calculated from
Eq. (\ref{TotalTN}). For both shot noise and technical noise it
can be shown that the noise of the total intensity is equal to sum
of the noise of each individual channel \cite{Lodahl04}. After
averaging over disorder, we obtain
\bs \bea && \left< \left< \mathcal{T}_a^{SN} \right> \right> =  \frac{\ell}{L}  , \label{T_SN} \\
&& \left< \left< \mathcal{T}_a^{TN} \right> \right> =
\frac{\ell^2}{L^2} \label{T_TN},
 \eea \label{TotalT}\es
where $\ell$ is the transport mean free path for multiple
scattering and $L$ the sample thickness. We have omitted
contributions from universal conductance fluctuations
\cite{Lodahl04,Feng88} that are negligible for the experiment
described in the following. In summary, we have predicted that the
total transmissions of quantum and classical noise vary linearly
and quadratically with the ratio of the mean free path to the
sample thickness, respectively.

Given the transmission coefficients of Eqs. (\ref{TN-SNonespot}),
we observe that the noise auto-correlation functions defined in
Eq. (\ref{corrfcts}) are either fourth-order or second-order
transmission correlation functions for technical noise and shot
noise, respectively. Assuming the electric field amplitudes can be
described by a circular Gaussian process  (short range
correlations) \cite{Goodman}, it follows that the fourth-order
correlation function can be expressed in terms of the second-order
correlation function, which implies
\bs \bea && \left< \left< C_{ab}^{SN}(\Delta \omega) \right> \right>_{\omega} = f(\eta), \\
&& \left< \left< C_{ab}^{TN}(\Delta \omega)\right>
\right>_{\omega} = f^2(\eta) + 4 f(\eta),  \eea \label{Icorr} \es
where $f(\eta) = \eta/\left[\cosh (\sqrt{\eta}) -
\cos(\sqrt{\eta}) \right]$ is the second-order correlation
function given in \cite{Berkovits94} with $\eta=2 L^2 \Delta
\omega/D$.

\begin{figure}[t]
 \includegraphics[width=\columnwidth]{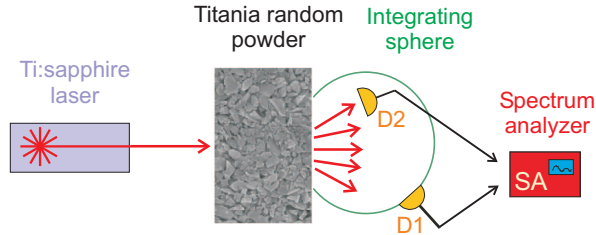}
 \caption{(color online). Experimental setup for measuring the transmission of quantum noise through a multiple scattering medium.
 Two different measurements were carried out by inserting either detector D1 or D2. The total transmission was
 recorded with an integrating sphere onto detector D1.
 With detector D2, the noise in a single speckle spot was measured. }
 \label{setup}
\end{figure}

The experimental setup is outlined in Fig. \ref{setup}. A
frequency tunable titanium-sapphire laser was used to probe the
random multiple scattering samples. The lasers' amplitude noise
spectrum was found to be limited by shot noise above $\sim1.5 \:
\mbox{MHz}$ and dominated by technical noise at lower frequencies,
which was carefully checked by observing that the noise of the
input beam scaled quadratically and linearly with intensity,
respectively. The two regimes enable us to study simultaneously
the transmission of classical and quantum noise. We used strongly
scattering samples consisting of titania particles (refractive
index $2.7$) with size distribution $d = 220 \pm 70 \: \mbox{nm}$
deposited on a fused silica substrate. Two types of experiments
were carried out: total transmission and speckle frequency
correlation measurements. In the former experiment, the
transmitted diffuse light was collected with an integrating sphere
onto a sensitive silicon photodiode (detector D1 in Fig.
\ref{setup}). The intensity noise was recorded by measuring the
spectral density of the photocurrent $\left|\delta i(\Omega)
\right|^2$ with a spectrum analyzer. Thermal noise from the
detector was subtracted in the measurements. The total
transmission of noise was obtained by dividing the measurements
with a noise spectrum recorded without any sample inserted. In the
speckle correlation measurements we recorded the intensity noise
in a single speckle spot (detector D2 in Fig. \ref{setup}) that
was selected using a pinhole. We subsequently varied the frequency
of the laser and recorded a frequency speckle pattern. In total
$200$ noise spectra were measured at equally spaced optical
frequencies with a frequency step of about $0.5 \: \mbox{THz}.$
From the complete measurement series the auto-correlation function
was obtained.

\begin{figure}[t]
\includegraphics[width=\columnwidth]{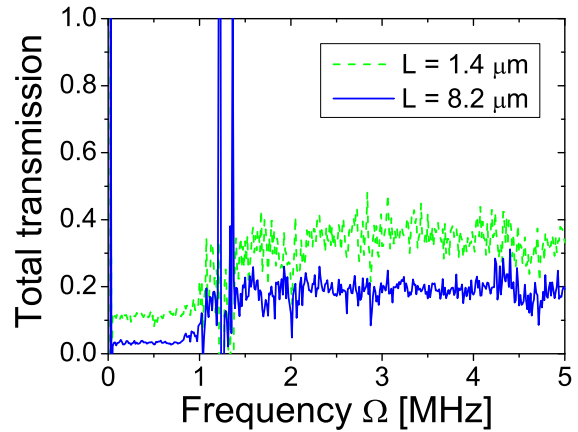}
 \caption{(color online). Total transmission of noise as a function of measurement frequency $\Omega$ for two different sample
 thicknesses. The spectral densities were recorded with a
 resolution bandwidth of $30 \: \mbox{kHz}$ and a video bandwidth
 of $10 \: \mbox{kHz}$ and by averaging each trace 100 times.
  Radically different transmissions are observed
 for technical noise (below $1 \: \mbox{MHz}$) compared to shot noise (above $1.5 \: \mbox{MHz}$). The spikes around $1.3 \: \mbox{MHz}$
 are due to oscillations in the detector power supply and are abandoned in the analysis.  }
 \label{rel_noise}
\end{figure}

Figure \ref{rel_noise} displays two measurements of the total
transmission of noise through samples with different thicknesses.
Two frequency regimes are apparent in the data: below $\sim 1 \:
\mbox{MHz}$ and above $\sim1.5 \: \mbox{MHz}$ corresponding to the
frequencies where the input laser light was dominated by technical
noise and shot noise, respectively. We observe immediately that
the total transmission of classical noise is significantly lower
than the total transmission of quantum noise. Within each noise
regime we average over the detection frequency $\Omega$, and
ensemble averaging is performed by measuring the transmission
several times at different positions on the sample.

\begin{figure}[t]
\includegraphics[width=\columnwidth]{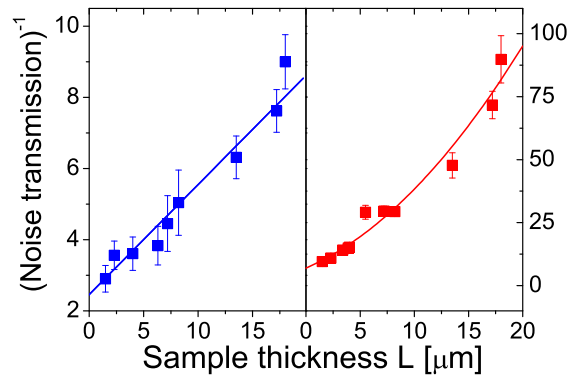}
 \caption{(color online). Left panel: inverse total transmission of quantum noise as a function of sample thickness.
 The line is a linear fit
 to the experimental data. Right panel: inverse total transmission
 of classical noise and a quadratic fit to the data. The different scales in the two plots clearly
 demonstrate that classical and quantum noise are transmitted
 differently.
  }
 \label{TT}
\end{figure}

Figure \ref{TT} shows the measured inverse total transmission as a
function of sample thickness for both shot noise and technical
noise.  The experimental data are modelled with the theory in Eqs.
(\ref{TotalT}), and very good agreement is observed. From the fits
we extract the transport mean free path, and derive $\ell = 1.19
\pm 0.33 \: \mu \mbox{m}$ from the shot noise measurements and
$\ell = 1.03 \pm 0.09 \: \mu \mbox{m}$ from the technical noise
measurements. The two values agree to within the error-bars of the
measurements. A proper account for the boundaries of the sample
has been included by effectively extending the sample thickness
with extrapolation lengths determined by Fresnel corrections
\cite{Zhu91}. The boundary contributions are determined from the
theoretical model at $L=0$, and full consistency between the two
data sets was obtained.

\begin{figure}[t]
\includegraphics[width=\columnwidth]{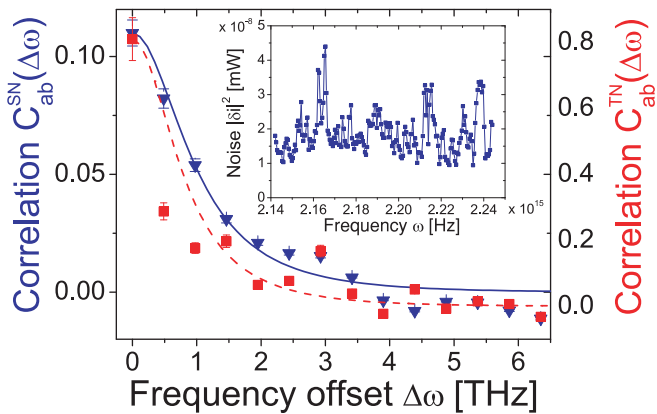}
\caption{(color online). Measured correlation function for shot
noise (triangles) and technical noise (squares) as a function of
frequency offset.
  The experimental data are compared to
  $\left< \left<C_{ab}^{SN} \right> \right>$ (full curve) and $
  \left< \left<C_{ab}^{TN} \right> \right>$ (dashed curve), respectively.
  Note the different magnitude of the correlation for quantum noise (left axis) and
  classical noise (right axis).
  The inset shows the complete shot noise data set of
the photocurrent spectral density $\left| \delta i \right|^2$ in a
speckle spot.  Each data-point was obtained by averaging the noise
spectra over the measurement frequency $\Omega$ within the limits
of the shot noise region. } \label{noise_versus_freq}
\end{figure}

In the speckle correlation measurements we again compare the
behavior of classical and quantum noise. The inset in Fig.
\ref{noise_versus_freq} displays the frequency speckle obtained
for shot noise by varying the optical frequency of the incident
light. We compute the auto-correlation functions defined in Eq.
(\ref{corrfcts}) for both technical noise and shot noise, and
their decay with frequency are shown in Fig.
\ref{noise_versus_freq}. The theoretical correlation functions
were corrected for a reduced contrast due to stray intensity
associated with the selection of a single speckle spot in the
experiment. The contrast can be well described by a constant
background for both data sets \cite{background}.
  The correlation function for shot noise is found to extend further
in frequency than the correlation function for technical noise,
which occurs since the former decays as second-order correlation
function and the latter as a fourth-order correlation function.
The measured noise correlation functions can be fitted well by the
theoretical prediction of Eqs. (\ref{Icorr}) using the known value
of the sample thickness ($L = 18 \: \mu \mbox{m}$). We derive the
diffusion constant $D = 34 \pm 2 \: \mbox{m}^2/\mbox{s}$ from the
shot noise data and $D = 30 \pm 4 \: \mbox{m}^2/\mbox{s}$ from the
technical noise data that both are consistent with time-resolved
propagation experiments on similar samples. The slight
oscillations in the experimental data (most clearly visible for
the technical noise data) could be a result of the limited
statistics of 200 measurement points corresponding to about 20
independent speckle spots. The good agreement between theory and
 experimental data for the quantum noise measurements confirms
the validity of the quantum model for multiple scattering.

As a side result our experiments demonstrate the robustness of
shot noise in multiple scattering: no excess noise was observed
due to scattering as opposed to what is expected for an amplifying
medium \cite{Patra99,Patra00}. This illustrates an important
difference between shot noise in electronics and optics. In a
disordered metal wire electronic shot noise is corrupted by
thermal noise on the scale of the electron-phonon scattering
length \cite{Blanter00}, and even for purely elastic electron
scattering is shot noise reduced by a factor of three
\cite{Beenakker92}. On the contrary, optical shot noise prevails
over distances much longer than the (elastic) scattering length.

We have studied the propagation of classical and quantum noise
through a multiple scattering medium. Both static and dynamic
measurements were carried out and compared to theory, which
allowed extracting fundamental scattering properties of the
medium. The quantum fluctuations were found to scale markedly
different with scattering parameters compared to classical
fluctuations, hence explicitly demonstrating the difference
between particle-like and wave-like transmission.

We acknowledge Allard Mosk, Ivo Vellekoop, and Valentin Freilikher
for valuable discussions. This work is part of the research
program of the "Stichting voor Fundamenteel Onderzoek der Materie
(FOM)", which is financially supported by the "Nederlandse
Organisatie voor Wetenschappelijk Onderzoek (NWO)".


\begin{thebibliography}{}
\bibitem{background}
The contrast is modeled by a single constant offset parameter
$\Delta$. Full consistency between the shot noise and technical
noise data was found since the relative offsets agree very well:
$\Delta^{SN}/\left< \left< \left|\delta I^{SN} \right|^2\right>
\right>_{\omega} = 2.0 $ and $\Delta^{TN}/\left< \left< \left|
\delta I^{TN} \right|^2\right> \right>_{\omega} = 2.3 .$
\end{thebibliography}
\end{document}